\documentstyle[twoside,aps]{revtex}
\newcommand{\snl}{\stackrel{(0)}{<}}
\newcommand{\snr}{\stackrel{(0)}{>}}
\begin{document}
\draft
\title{Optimized Rayleigh-Schr\"{o}dinger Expansion \\
        of the Effective Potential}
\author{Wen-Fa Lu$^{a,b,c}$, Chul Koo Kim$^{a,c}$, and Kyun Nahm$^d$}
\address{\begin{flushleft}
\hspace*{2.2cm}$^a$ Institute of Physics and Applied Physics, Yonsei
University, Seoul 120-749, Korea \\
\hspace*{2cm} $^b$ Department of Physics, Shanghai Jiao Tong University,
Shanghai 200030,\\ \hspace*{2.2cm} the People's Republic of China
   \thanks{permanent address,E-mail: wenfalu@online.sh.cn} \\
 \hspace*{2.cm}$^c$ Center for Strongly Correlated Materials Research,
Seoul National University, \\ \hspace*{2.2cm}Seoul 151-742, Korea \\
\hspace*{1.8cm} $^d$ Department of Physics, Yonsei University, Wonju 220-710,
Korea
     \end{flushleft}}
%\date{}
\maketitle

\begin{abstract}
An optimized Rayleigh-Schr\"{o}dinger expansion scheme of solving the
functional Schr\"odinger equation with an external source is proposed to
calculate the effective potential beyond the Gaussian approximation. For a
scalar field theory whose potential function has a Fourier representation in a
sense of tempered distributions, we obtain the effective potential up to the
second order, and show that the first-order result is just the Gaussian
effective potential. Its application to the $\lambda\phi^4$ field theory yields
the same post-Gaussian effective potential as obtained in the functional
integral formalism.
\end{abstract}
%\vspace{24pt}
Author Keywords: Effective potential; A Class of scalar field
models;Functional Schr\"{o}dinger equation; Variational
perturbation approach; non-perturbative quantum field theory

PACS classification codes: 11.10.-z; 11.10.Lm; 11.15.Tk

\section{Introduction}
\label{1}

Since the mid 1980s, the effective potential (EP) in the quantum
field theory (QFT) \cite{1} beyond the Gaussian approximation (GA)
has received much attention \cite{2}, because it collects merits
and removes weakpoints of the conventional perturbation theory and
the GA as shown by the so-called variational perturbation theory
in other fields \cite{3}. Some schemes have proposed to calculate
such an EP, for example, designing some non-Gaussian trial
wavefunctionals \cite{2}(i), using the Brueckner-Goldstone formula
with a variational basis \cite{2}(ii), rearranging loop diagrams
in the functional integral formalism with the background field
method \cite{2}(iii), optimized expansions in the functional
integral formalism with the steepest-descent method \cite{2}(iv)
and with the background field method \cite{2}(v), and so on.
Noting that the well-developed Rayleigh-Schr\"{o}dinger (RS)
expansion in quantum mechanics \cite{4} was generalized to QFT
\cite{5}, recently we have developed a variational perturbation
scheme with the RS expansion to calculate the EP as per the
variational minimum definition \cite{6}. In the scheme, a
free-field Hamiltonian with a mass fixed from the GA was adopted
as a solvable part so as to perform the RS expansion \cite{6}.
Obviously, this scheme amounts to a series expansion around the
Gaussian EP, and so do those schemes in Ref.~\cite{2} (i)---(iii).
In the present paper, taking a free-field Hamiltonian instead with
an arbitrary mass parameter $\mu$ as the solvable part, we will
apply the RS expansion to solve the functional Schr\"{o}dinger
equation in presence of an external source so that the vacuum
energy functional of the source can be obtained. Then, adopting
the alternative, but equivalent definition of the EP in the
functional Schr\"{o}dinger picture \cite{5}, we extract the EP
through a Legendre transform of the energy functional. In the new
scheme, $\mu$ will be determined according to the principle of
minimal sensitivity \cite{7} and, consequently, the value of $\mu$
in an approximation up to one order is different from those up to
other orders of the expansion. This way of fixing $\mu$ makes the
expansion give the Gaussian EP with its lowest orders, but be not
an expansion around the Gaussian EP. Furthermore, unlike
Ref.~\cite{6}, the vacuum expectation value $\varphi$ of the field
operator will naturally be given in the present scheme. Hereafter,
we will call this scheme the optimized RS expansion (ORSE). Since,
for the case of scalar field theory, the EP beyond the GA was
given only for the $\lambda \phi^4$ model except for the $\phi^6$
model \cite{2}(1985) up to now, we use the ORSE in this paper to
give the EP for a generic class of scalar field models (see Sect.
III) up to the second order. The resultant formula can easily be
used to give the EP beyond the GA for a number of concrete scalar
field models, for instance, models with polynomial or/and
exponential interactions. We also show that its application to
$\lambda\phi^4$ field theory gives the same post-Gaussian EP as
Ref.~\cite{2}(iv,v), and yields the result in
Ref.~\cite{6}\cite{2}(ii,iii) if one chooses to use the same
constraint on $\varphi$ as in Ref.~\cite{2}(ii).

Next, for the sake of convenience, we will first give the complete basis set
for a free field theory with an external source, which will be employed in
the ORSE. In Sect. III, the ORSE will be proposed. In Sect. IV, we will
perform the ORSE to obtain the EP for a class of scalar field theories up to
the second order, and its application to the $\lambda\phi^4$ field theory will
be given in Sect. V. Conclusions will be made in Sect. VI with discussions on
some possible extensions and developments of the present work.

\section{A Free Field Theory with an External Source}
\label{2}

In this section, we discuss the free-field theory with an external source
$J_x\equiv J(\vec{x})$ in a time-fixed functional Schr\"{o}dinger picture. The
Hamiltonian is given by
\begin{equation}
H^{J,\mu}_0=\int_x [ {\frac {1}{2}}\Pi_x^2
                +{\frac {1}{2}}(\partial_x \phi_x)^2 +
                {\frac {1}{2}}\mu^2 \phi_x^2 -J_x\phi_x-{\frac {1}{2}} f_{xx}
                +{\frac {1}{2}}\int_y J_x h_{xy}^{-1} J_y] ,
\end{equation}
where $x=(x^1,x^2, \cdots, x^D)$ represents a position in $D$-dimensional
space, $\int_x\equiv \int d^D x$, $\mu$ an arbitrary mass parameter and
$\phi_x\equiv \phi(\vec{x})$ the field at $x$. $\Pi_x\equiv -i{\frac
{\delta}{\delta \phi_x}}$ is canonically conjugate to $\phi_x$
with the commutation relation, $[\phi_x,\Pi_y]=i\delta(x-y)$. In Eq.(1),
$f_{xy}\equiv (\sqrt{-\partial_x^2 +\mu^2})\delta(x-y)$ with $\int_z
f_{xz}f_{zy}^{-1}=\delta(x-y)$, and $h_{xy} \equiv (-\partial_x^2 +\mu^2)
\delta(x-y)$ with $\int_z h_{xz}h_{zy}^{-1}= \delta(x-y)$.

The functional Schr\"{o}dinger equation for Eq.(1), $H^{J,\mu}_0|n;J\snr=
E_n^{(0)}[J] |n;J\snr$, is easily solved \cite{5} (Here, the subscript $n$ in
$E_n^{(0)}[J]$ is the index of eigenstates, and the superscript $``(k)''$ means
``at the $k$th order of $\delta$''. See the next section.). First, the
eigenenergy of the vacuum state vanishes, and the corresponding wavefunctional
is a Gaussian-type functional
\begin{equation}
|0;J\snr = {\cal N} \exp\{ -{\frac {1}{2}}\int_{x,y}
    (\phi_x-\int_z h_{xz}^{-1} J_z)f_{xy}(\phi_y-\int_z h_{yz}^{-1} J_z)\} \;,
\end{equation}
where ${\cal N}$ is the normalization constant ($i.e., \snl J;0|0;J\snr=1$).
One can show that $\snl J;0|\phi_x|0;J\snr=\int_z h_{xz}^{-1} J_z$ which,
unlike Eq.(3) in Ref.~\cite{6}, is dependent on $x$. Then, for the above vacuum,
the annihilation and creation operators can respectively be constructed as
\begin{equation}
A_f(p;J) = ({\frac {1}{2(2\pi)^D f(p)}})^{1/2}\int_x e^{-ipx}
   [f(p)(\phi_x - \int_z h_{xz}^{-1} J_z) + i\Pi_x]
\end{equation}
and
\begin{equation}
A^\dagger_f(p;J) = ({\frac {1}{2(2\pi)^D f(p)}})^{1/2}\int_x e^{ipx}
   [f(p)(\phi_x - \int_z h_{xz}^{-1} J_z) - i\Pi_x]
\end{equation}
with $[A_f(p;J), A^\dagger_f(p';J)]=\delta (p'-p)$ and $A_f(p)|0\snr=0$. It is
not difficult to verify that $H^{J,\mu}_0=\int d^D p f(p) A^\dagger_f(p;J)
A_f(p;J)$, where $f(p)=\sqrt{p^2+\mu^2}$ arises from $f_{xy}=\int d^D p
f(p)e^{i p(x-y)}$ with $p=(p^1, p^2, \cdots, p^D)$. Consequently, the
eigenwavefunctionals for excited states can be easily written as
\begin{equation}
|n;J\snr ={\frac {1}{\sqrt{n!}}} \prod_{i=1}^n A^\dagger_f(p_i;J)|0;J\snr
          ,  \; \; n= 1, 2, \cdots, \infty
\end{equation}
and the corresponding eigenenergies are
\begin{equation}
E^{(0)}_n[J]=\sum_{i=1}^n f(p_i) .
\end{equation}
Evidently, the eigenwavefunctionals $|n;J\snr$ and $|0;J\snr$ are orthogonal
and normalized, $\snl J;m|n;J\snr =\delta_{mn}{\frac {1}{n!}} \sum_{P_i(n)}
\prod_{k=1}^n\delta (p'_{k}-p_{i_k})$. Here, $P_i(n)$ represents a permutation
of the set $\{i_k\}=\{1,2,\cdots,n\}$ and the summation is over all
$P_i(n)s$. $|n;J\snr$ describes a $n$-particle state with the continuous
momenta $p_1, p_2,\cdots, p_n$. $|0;J\snr$ and $|n;J\snr$ with $n=1, 2, \cdots,
\infty$ constitute the complete set for $H^{J,\mu}_0$, and satisfy the closure
$|0;J\snr\snl J;0|+\sum_{n=1}^\infty \int d^D p_1d^D p_2\cdots d^D p_n|n;J\snr
\snl J;n|=1$.

\section{Optimized Rayleigh-Schr\"{o}dinger Expansion \\
       for the Effective Potential}
\label{3}

The EP for a field system is equivalently defined through the Feynman graphs,
the operator representation, the path integral \cite{1}, or the minimum
expected energy in a set of normalized states \cite{5,8}. They were used to give
the loop or Gaussian EP and propose those schemes in Refs.~\cite{2,6}. Yet
another equivalent definition of the EP is given through the vacuum energy
functional of an external source obtained by solving the relevant functional
Schr\"{o}dinger equation \cite{5}. Based on it, we will construct the ORSE in
this section.

In this and next sections, we work with a scalar field model whose Lagrangian
density is \cite{9}
\begin{equation}
{\cal L}={\frac {1}{2}}\partial_\mu \phi_x \partial^\mu \phi_x
-V(\phi_x) \;.
\end{equation}
In Eq.(7), the model potential is assumed to be written as
$V(\phi_x)=\int {\frac {d \Omega}{\sqrt{2\pi}}} \tilde{V}(\Omega)
e^{i\Omega\phi_x}$, at least, in a sense of tempered distributions \cite{10}.
It represents several scalar-field models, such as $\lambda\phi^4$ model
\cite{2,11}, general and special $\phi^6$ models \cite{12}, sine-Gordon and
sinh-Gordon models \cite{13}, massive and double sine-Gordon model \cite{14},
Liouville model \cite{15}, as well as two generic models investigated in
Ref.~\cite{16}.

For the system, Eq.(7), the time-independent functional Schr\"{o}dinger
equation in the presence of an external source $J_x$ is
\begin{equation}
(H-\int_x J_x \phi_x)|\Psi_n>=E_n[J]|\Psi_n>
\end{equation}
with the Hamiltonian $H=\int_x [{\frac {1}{2}}\Pi_x^2+{\frac {1}{2}}
(\partial_x \phi_x)^2+V(\phi_x)] $.
Here, the eigenvalue $E_n[J]$ is a functional of $J_x$.

For our purpose, Eq.(8) will be modified. We make a shift $\phi_x\to
\phi_x+\Phi$ ($\Phi$ is a constant), and Eq.(8) can equivalently be rewritten
as
\begin{equation}
[H(\phi_x+\Phi)-\int_x J_x (\phi_x+\Phi)]\Psi_n[\phi_x+\Phi,J]
=E_n[J;\Phi]\Psi_n[\phi_x+\Phi,J]
\end{equation}
with $H(\phi_x+\Phi)=\int_x [{\frac {1}{2}}\Pi_x^2+{\frac {1}{2}}(\partial_x
      \phi_x)^2+V(\phi_x+\Phi)]$.
This shift is really in the spirit of the background-field method \cite{17}
\cite{2}(iii). Further, normal-ordering the Hamiltonian in Eq.(9) with respect
to a normal-ordering mass $M$ \cite{18}, and inserting a vanishing term $\int_x
[{\frac {1}{2}}\mu^2 \phi_x^2-{\frac {1}{2}}\mu^2 \phi_x^2]$ with $\mu$ an
arbitrary mass parameter into the Hamiltonian\cite{19}, one can have
\begin{equation}
{\cal N}_M[H(\phi_x+\Phi)-\int_x J_x (\phi_x+\Phi)]=
       H_0^{J,\mu}+H_I^{\mu,\Phi}-C
\end{equation}
with
\begin{equation}
H_I^{\mu,\Phi}=\int_x\{-{\frac {1}{2}}\mu^2 \phi_x^2+
{\cal N}_M[V(\phi_x+\Phi)]\}
\end{equation}
and
\begin{equation}
C=\int_x [-{\frac {1}{2}} f_{xx}
  +{\frac {1}{2}}\int_y J_x h_{xy}^{-1} J_y+{\frac {1}{2}}I_0(M^2)
  -{\frac {M^2}{4}}I_1(M^2)+J_x\Phi] \;.
\end{equation}
Here, the notation ${\cal N}_M[\cdots]$ represents normal-ordered form with
respect to $M$, $I_n(Q^2)\equiv \int {\frac {d^D p} {(2\pi)^D}}{\frac
{\sqrt{p^2+Q^2}}{(p^2+Q^2)^n}}$, and ${\cal N}_M[V(\phi_x+\Phi)]=\int
{\frac {d \Omega}{\sqrt{2\pi}}} \tilde{V}(\Omega)e^{i\Omega(\phi_x+\Phi)
+{\frac {M^2}{4}}I_1(M^2)} $ \cite{9}. From now on, we will use the
normal-ordered Hamiltonian in Eq.(10) instead of the original one, which will
naturally make the EP in (1+1) dimensions free of explicit ultraviolet
divergences \cite{9,18}. Noting that $C$ is a constant independent of $\phi_x$
and $H_0^{J,\mu}$ is an exactly-solved Hamiltonian, we can formally treat
$H_I^{\mu,\Phi}$ as a ``perturbed'' interaction in the RS expansion \cite{4}.
To mark the order of the RS expansion, an index factor $\delta$ will be
inserted in front of $H_I^{\mu,\Phi}$ in Eq.(10). Consequently, Eq.(9) is
modified as
\begin{equation}
[H_0^{J,\mu}+\delta H_I^{\mu,\Phi}]\Psi_n[\phi_x+\Phi,J;\delta]
         =(E_n[J;\Phi,\delta]+C)\Psi_n[\phi_x+\Phi,J;\delta] \;.
\end{equation}
Now, applying the RS expansion, one can solve Eq.(13) to get energy eigenvalues,
$E_n[J;\Phi,\delta]$, and eigenwavefunctionals, $\Psi_n[\phi_x+\Phi,J;\delta]$.
Here, we are interested only in the eigenenergy functional for vacuum state,
$E_0[J;\Phi,\delta]$. Obviously, the zeroth-order approximation to
$E_0[J;\Phi,\delta]$, $E_n^{(0)}[J;\phi]$, satisfies
\begin{equation}
E_0^{(0)}[J;\phi]+C=E^{(0)}_0[J]=0
\end{equation}
and at the $n$th-order of $\delta$, the correction to
$E_0^{(0)}[J;\phi]$ is
\begin{equation}
E_0^{(n)}[J;\phi]=\snl 0|H_I^{\mu,\Phi}
        [Q_0 {\frac {1}{H_0^{J,\mu}-E_n^{(0)}[J]}}(E_n^{(1)}[J;\phi]
        -H_I^{\mu,\Phi})]^{n-1}|0\snr
\end{equation}
with $Q_n=\sum_{j\not=n}^\infty \int d^D p_1d^D p_2\cdots d^D p_j|j\snr
\snl j|$. Thus, $E_0[J;\Phi,\delta]=E_0^{(0)}[J;\phi]
+\sum_{n=1}^{\infty}\delta^n E_0^{(n)}[J;\phi]$.

With the vacuum energy functional $E_0[J;\Phi,\delta]$, one can have
\begin{equation}
{\frac {\delta E_0[J;\Phi,\delta]}{\delta J_x}}
=-\int {\cal D}\phi \Psi^\dagger_0[\phi_x+\Phi,J;\delta]
(\phi_x+\Phi) \Psi_0[\phi_x+\Phi,J;\delta]\equiv -\varphi_x
\end{equation}
where we used the Feynman-Hellmann theorem \cite{20}. Evidently, for $\delta=1$,
$\varphi_x=<\Psi_0|\phi_x|\Psi_0>$. Then, a Legendre
transformation of $E_0[J;\Phi,\delta]$ yields the static effective action
\cite{5}
\begin{equation}
\Gamma_s[\varphi;\Phi,\delta]=-E_0[J;\Phi,\delta]-\int_x J_x \varphi_x \;.
\end{equation}
To calculate EP, one can conveniently take $\varphi_x=\Phi$ in Eq.(16) to fix
the arbitrary shifted parameter $\Phi$. In analogy to Appendix A of
Ref.~\cite{2}(1990) but with the Feynman-Hellmann theorem \cite{20}, one can
show that other choices of $\Phi$ will give rise to the same EP (when the
wavefunction renormalization procedure is not needed). Finally, one can have
the EP \cite{5}
\begin{equation}
{\cal V}(\Phi)\equiv -{\frac {\Gamma_s[\Phi,\delta;\varphi]}{\int_x}}
                    \biggl |_{\varphi_x= {\rm constant}=\Phi,
                    \delta=1} \;.
\end{equation}
If the EP is truncated at some order of $\delta$, the extrapolation
of $\delta=1$ should be made after renormalizing the approximated EP.

When truncated at a given order of $\delta$, the right side of Eq.(18) will
depend upon $\mu$. To obtain an approximated EP up to the same order, one can
determine $\mu$ according to the principle of minimal sensitivity \cite{7}. That
is, $\mu$ should be such a value that the approximated EP up to the given order
is optimized to be as insensitive to variations in $\mu$ as possible. This can
be realized by analyzing the vanishing first- or higher-order derivatives
of the truncated result with respect to $\mu$ \cite{7}\cite{2}(v). From the next
section, one will see that the EP up to the first order is just the Gaussian EP.
Thus, the ORSE is a systematic tool of improving the Gaussian EP.

\section{Optimized Effective Potential for a Class of Models \\
         up to the Second Order}
\label{4}

In this section, we carry out the ORSE for the system, Eq.(7), to
calculate the EP up to the second order.

The matrix elements which appear in Eq.(15) involve only Gaussian integrals
except commutators of creation and annihilation operators and, thus, can be
readily calculated as follows,
\begin{eqnarray}
\snl n|&{\cal N}_M[V(\phi_x+\Phi)]&|m\snr = {\frac{1}{\sqrt{n!m!}}}
     \sum_{i=0}^n C_n^i C_m^{m-n+i} (n-i)!
     (2(2\pi)^{D})^{-{\frac {m-n+2i}{2}}}
     \nonumber \\ &\ \ \ &
     \cdot (\prod_{j=n-i+1}^{n}f(p_j)
     \prod_{k=n-i+1}^{m}f(p'_k))^{-{\frac {1}{2}}}
     e^{i(\sum_{k=n-i+1}^{m}p'_k-\sum_{j=n-i+1}^{n}p_j)x}
    \prod_{l=1}^{n-i}\delta(p'_l-p_l) \nonumber \\ &&
  \cdot \int_{-\infty}^{\infty}{\frac {d \alpha}{\sqrt{\pi}}} e^{-\alpha^2}
    V^{(m-n+2i)}(\alpha \sqrt{f_{xx}^{-1}-I_1(M^2)}+\Phi+\int_z
        h_{xz}^{-1}J_z)
\end{eqnarray}
with $n\leq m$.
In Eq.(19), $V^{k}(z)\equiv {\frac {d^k V(z)}{(dz)^k}}$. For simplicity, in
getting the above results, we have employed the permutation symmetry of momenta
in Eq.(15) for various products of $\delta$ functions. Note that matrix elements
of $\phi_x^2$ are special cases of Eq.(19).

Substituting the above matrix elements into Eq.(15), one can obtain the first-
and the second-order corrections to $E_0^{(0)}[J;\phi]$ as
\begin{eqnarray}
E_0^{(1)}[J;\phi]&=&
    \int_x\{-{\frac {\mu^2}{2}}[(\int_z
       h_{xz}^{-1}J_z)^2+{\frac {1}{2}}f_{xx}^{-1}]
         \nonumber  \\   &\ \ \ &
       + \int_{-\infty}^{\infty}{\frac {d \alpha}{\sqrt{\pi}}} e^{-\alpha^2}
    V(\alpha \sqrt{f_{xx}^{-1}-I_1(M^2)}+\Phi+\int_z
        h_{xz}^{-1}J_z)\}
\end{eqnarray}
and
\begin{eqnarray}
E_0^{(2)}[J;\phi]&=&
   -{\frac {\mu^4}{2}}\int {\frac {d^D p}{(2\pi)^D}}{\frac {1}{f^2(p)}}\biggl
   |\int_{xz} e^{ipx} h_{xz}^{-1}J_z\biggl |^2 -
   {\frac {\mu^4}{16}}\int {\frac {d^D p}{(2\pi)^D}}{\frac {1}{f^3(p)}}\int_x
               \nonumber \\ &\ \ \ &
   +\mu^2\int {\frac {d^D p}{(2\pi)^D}}{\frac {1}{f^2(p)}}
   \int_{x_1 z} e^{-ipx_1} h_{x_1 z}^{-1}J_z\int_{x_2} e^{ipx_2}
   \int_{-\infty}^{\infty}{\frac {d \alpha}{\sqrt{\pi}}} e^{-\alpha^2}
               \nonumber \\ &\ \ \ \ \ \ \ \ &
 \ \ \ \ \ \cdot V^{(1)}(\alpha \sqrt{f_{xx}^{-1}-I_1(M^2)}+\Phi+\int_z
        h_{xz}^{-1}J_z)
       \nonumber \\ &\ \ \ &
   +{\frac {\mu^2}{8}}\int {\frac {d^D p}{(2\pi)^D}}{\frac {1}{f^3(p)}}
   \int_x \int_{-\infty}^{\infty}{\frac {d \alpha}{\sqrt{\pi}}} e^{-\alpha^2}
    V^{(2)}(\alpha \sqrt{f_{xx}^{-1}-I_1(M^2)}+\Phi+\int_z
        h_{xz}^{-1}J_z)
          \nonumber \\ &\ \ \ &
    -\sum_{j\not=0}^{\infty} {\frac {1}{j!2^j}}
    \int {\frac {\prod_{k=1}^j d^D p_k}{(2\pi)^{jD}}}
    {\frac {1}{\prod_{k=1}^j f(p_k) (\sum_{k=1}^j f(p_k))}}
               \nonumber \\ &\ \ \ \ \ \ \ \ &
   \cdot \biggl |\int_x e^{ix\sum_{k=1}^j p_j}
      \int_{-\infty}^{\infty}{\frac {d \alpha}{\sqrt{\pi}}} e^{-\alpha^2}
    V^{(j)}(\alpha \sqrt{f_{xx}^{-1}-I_1(M^2)}+\Phi+\int_z
        h_{xz}^{-1}J_z)\biggl |^2   \;,
\end{eqnarray}
respectively. Here, $``\biggl |\cdots \biggl |$'' represents the absolute
value.

Next, we extract the approximated EP for the system order by order.

At the zeroth order, $E_0^{(0)}[J;\phi]=-C$ from Eq.(14), and so, taking
$-{\frac {\delta E_0^{(0)}[J;\phi]}{\delta J_x}}=\int_y
h_{xy}^{-1}J_y+\Phi=\varphi_x^{(0)}$ as $\Phi$, one has $J^{(0)}=0$.
Consequently, the EP at the zeroth-order of $\delta$ is
\begin{equation}
{\cal V}^{(0)}(\Phi) =-{\frac {\Gamma_s^{(0)}[\varphi;\Phi,\delta]}{\int_x}}
                    \biggl |_{\varphi_x= \Phi} =
                    {\frac {1}{2}} f_{xx}
  -{\frac {1}{2}}I_0(M^2)
  +{\frac {M^2}{4}}I_1(M^2)  \;.
\end{equation}

Up to the first order (hereafter, any Greek-number superscript, such as $``I",
``II"$, means ``up to the order whose number is consistent with the Greek
number"),
\begin{equation}
E_0^{I}[\Phi,\delta;J]=E_0^{0}[J;\phi]+\delta E_0^{(1)}[J;\phi]
\end{equation}
and $-{\frac {\delta E_0^{I}[\Phi,\delta;J]}{\delta J_x}}=
\varphi_x^{I}=\Phi$ yields
\begin{eqnarray}
&&\int_y h_{xy}^{-1}J_y +\delta \int_y h_{xy}^{-1}\{\mu^2\int_z h_{yz}^{-1}J_z
   \nonumber  \\  &&  \ \ \ \ \ \ \ \ \ \
   -\int_{-\infty}^{\infty}{\frac {d \alpha}{\sqrt{\pi}}} e^{-\alpha^2}
    V^{(1)}(\alpha \sqrt{f_{xx}^{-1}-I_1(M^2)}+\Phi+\int_z
        h_{xz}^{-1}J_z)\}=0     \;.
\end{eqnarray}
When extracting the EP up to first order, Eqs.(17), (18) and (23) imply
that only the $J_x$ up to the first order, $J^{I}$, is necessary.
Owing to $J^{(0)}=0$, it is enough to take $J_x=0$ for the last term in the
left hand of Eq.(24). Thus, $J^{I}$ can be solved from Eq.(24) as
\begin{equation}
J^{I}=\delta \int_{-\infty}^{\infty}{\frac {d \alpha}{\sqrt{\pi}}}
    e^{-\alpha^2} V^{(1)}(\alpha \sqrt{f_{xx}^{-1}-I_1(M^2)}+\Phi) \;.
\end{equation}
Even $J^I$ will not be needed to get the EP up to the first order of $\delta$,
because there exists no linear, but the quadratic term of $J_x$ in the
zeroth-order term of $(E_0^{I}[J;\Phi,\delta]-\int_x J_x \Phi)$ as shown in
Eq.(12). In fact, to obtain the EP up to the $n$th order, one need the
approximated $J$ only up to the $(n-1)$th order. Now one can write down the EP
up to the first order
\begin{eqnarray}
{\cal V}^{I}(\Phi,\delta) &=&
                {\frac {1}{2}} [f_{xx} -I_0(M^2)] +{\frac {1}{4}}M^2I_1(M^2)
                    -\delta {\frac {1}{4}}\mu^2f_{xx}^{-1}
                    \nonumber \\  &\ \ \ &
     +\delta \int_{-\infty}^{\infty}{\frac {d \alpha}{\sqrt{\pi}}}
    e^{-\alpha^2} V(\alpha \sqrt{f_{xx}^{-1}-I_1(M^2)}+\Phi) \;.
\end{eqnarray}
Obviously, this result will yield nothing but the Gaussian EP \cite{9}
(1995,2002).

Finally, we consider the second order. $\varphi_x=\varphi_x^{II}=-{\frac
{\delta E_0^{II}[J;\Phi,\delta]}{\delta J_x}}=\Phi$ can be solved for $J^{II}$.
In the present case, however, it is enough to use only $J^{I}$ for the EP.
Substituting $J^I$ into Eq.(18), we obtain the EP for the system, Eq.(7), up
to the second order as
\begin{eqnarray}
{\cal V}^{II}(\Phi,\delta) &=&
              {\frac {1}{2}} [f_{xx} -I_0(M^2)]+{\frac {1}{4}}M^2I_1(M^2)
                    -\delta {\frac {1}{4}}\mu^2f_{xx}^{-1}
                    \nonumber \\  &\ \ \ &
       +\delta \int_{-\infty}^{\infty}{\frac {d \alpha}{\sqrt{\pi}}}
    e^{-\alpha^2} V(\alpha \sqrt{f_{xx}^{-1}-I_1(M^2)}+\Phi)
        \nonumber \\  &\ \ \ &
 -\delta^2 {\frac {\mu^2}{16}}\int {\frac {d^D p}{(2\pi)^D}}{\frac {1}{f^3(p)}}
   [\mu^2-2\int_{-\infty}^{\infty}{\frac {d \alpha}{\sqrt{\pi}}} e^{-\alpha^2}
    V^{(2)}(\alpha \sqrt{f_{xx}^{-1}-I_1(M^2)}+\Phi)]
        \nonumber \\  &\ \ \ &
    - \delta^2\sum_{j=2}^{\infty} {\frac {1}{j!2^j}}
    \int {\frac {\prod_{k=1}^{j-1} d^D p_k}{(2\pi)^{(j-1)D}}}
    {\frac {1}{f(\sum_{k=1}^{j-1} p_k)\prod_{k=1}^{j-1} f(p_k)}}
        \nonumber \\  &\ \ \ &
  \cdot  {\frac {1}{(f(\sum_{k=1}^{j-1} p_k)+\sum_{k=1}^{j-1} f(p_k))}}
 \biggl [\int_{-\infty}^{\infty}{\frac {d \alpha}{\sqrt{\pi}}} e^{-\alpha^2}
    V^{(j)}(\alpha \sqrt{f_{xx}^{-1}-I_1(M^2)}+\Phi)\biggl ]^2 \;,
\end{eqnarray}
where, one should take $\delta=1$ after renormalizing ${\cal V}^{II}
(\Phi,\delta)$, and $\mu$ is determined from the stationary condition
\begin{equation}
{\frac {\partial {\cal V}^{II} (\Phi)} {\partial \mu}} =0 \;.
\end{equation}
Here, ${\cal V}^{II} (\Phi)$ is the EP after ${\cal V}^{II} (\Phi,\delta)$ is
renormalized. If Eq.(28) has no real solutions, $\mu$ can be fixed by ${\frac
{\partial^2 {\cal V}^{II} (\Phi)} {(\partial \mu)^2}}=0$ \cite{7}.
Note that in (1+1) dimensions, $\{{\frac {1}{2}} [f_{xx} -I_0(M^2)]
+{\frac {1}{4}}M^2I_1(M^2) - {\frac {1}{4}}\mu^2f_{xx}^{-1}\}$ and $[f_{xx}^{-1}
-I_1(M^2)]$ in Eq.(27) with $\delta=1$ is finite and, thus, for any
(1+1)-dimensional theories which make the series in Eq.(30) finite, no
renormalization procedure is needed.

Similarly, employing Eq.(19), one can obtain higher order corrections to the
Gaussian EP from Eq.(15). To conclude this section, we emphasize that Eq.(27)
can easily be used to give the EPs for a number of scalar field theories
including those discussed in Refs.~\cite{2,11,12,13,14,15,15}.

\section{Application to $\lambda\phi^4$ field theory}
\label{5}

In this section, we consider the potential,
\begin{equation}
V(\phi_x)={\frac {1}{2}}m^2\phi_x^2+\lambda\phi_x^4  \;,
\end{equation}
which was widely studied with several variational perturbation
techniques \cite{2,6}.

Substituting Eq.(29) into Eq.(27), and noting that $\int_{-\infty}^{\infty}
\alpha^{2n+1}e^{-\alpha^2}{\frac {d \alpha}{\sqrt{\pi}}}=0$ for $n=1,2,\cdots$
and $\int_{-\infty}^{\infty} \alpha^{2n} e^{-\alpha^2}{\frac {d \alpha}
{\sqrt{\pi}}}=2^{-n}\cdot 1\cdot3\cdots (2n-1)$ for $n=0,1,2,\cdots$, one can
easily obtain the EP for the system, Eq.(29), up to the second order
\begin{eqnarray}
{\cal V}^{II}(\Phi,\delta)&=& {\frac {1}{2}} (f_{xx} -I_0(M^2))
             +{\frac {1}{4}}M^2I_1(M^2)+
         \delta({\frac {1}{2}}m^2\Phi^2+\lambda\Phi^4
         -{\frac {1}{4}}\mu^2f_{xx}^{-1}) \nonumber \\  &\ \ \ &
        +\delta{\frac {1}{4}}(f_{xx}^{-1}-I_1(M^2))
        [m^2+12\lambda\Phi^2+3\lambda (f_{xx}^{-1}-I_1(M^2))]
        \nonumber \\  &\ \ \ &
   -\delta^2{\frac {1}{16}}\int {\frac {d^D p}{(2\pi)^D}}{\frac {1}{f^3(p)}}
   [m^2-\mu^2+12\lambda\Phi^2+6\lambda (f_{xx}^{-1}-I_1(M^2))]^2
        \nonumber \\  &\ \ \ &
   -\delta^2 12\lambda^2\Phi^2\int {\frac {d^D p_1 d^D p_2}{(2\pi)^{2D}}}
    {\frac {1}{f(p_1)+f(p_2)+f(p_1+p_2)}}{\frac {1}{f(p_1)f(p_2)f(p_1+p_2)}}
        \nonumber \\  &\ \ \ &
   -\delta^2{\frac {3}{2}} \lambda^2\int {\frac {d^D p_1 d^D p_2 d^D p_3}
   {(2\pi)^{3D}}}{\frac {1}{f(\sum_{k=1}^{3} p_k)+\sum_{k=1}^{3} f(p_k)}}
   {\frac {1} {f(\sum_{k=1}^{3} p_k)\prod_{k=1}^{3} f(p_k)}} \;.
\end{eqnarray}
Discarding terms with $I_n(M^2) (n=0,1)$, the above result becomes identical to
Eq.(2.36) in Ref.~\cite{2}(1990). This can be verified by carrying out
integrations of $I_n(\Omega) (n=0,1)$ and $I^{(n)}(\Omega) (n=2,3,4)$ in
Ref.~\cite{2}(v) over one component of each Euclidean momentum. In $(1+1)$
dimensions, taking $\delta=1$, one finds that Eq.(30) is finite. Because the
(1+1)-dimensional Gaussian EP in the Coleman's normal-ordering prescription is
consistent \cite{9} (2002) with that in the Stevenson's reparametrization
scheme \cite{11} (1985), Eq.(30) with $\delta=1$ and $D=1$ is consistent with
the $\Delta m_B^2=0$ version of Sect. IV in Ref.~\cite{2} (1991). As for the
case of $(2+1)$ dimensions, Stancu has performed a renormalization procedure to
make ${\cal V}^{II}(\Phi)$ finite \cite{2} (1991).

Using $\mu$ fixed at the GA result for each order of the ORSE will
simultaneously imply that up to each order, the vacuum expectation value of the
field operator $\phi_x$ is identical to that in the GA. If one chooses to do
so, Eq.(30) and also the second-order result in our former work Ref.~\cite{6}
will yield the result in Ref.~\cite{2}(ii,iii).

Finally, we point out that taking $\mu=m$ will lead to the conventional
perturbation result on the EP.

\section{Discussion and Conclusion}
\label{6}

In this paper, an optimized RS expansion scheme of solving the functional
Schr\"{o}dinger equation with an external source is proposed to calculate the
EP beyond the GA. For the class of scalar field theories, Eq.(7), we obtain a
general expression on the EP up to the second order which can be used easily to
obtain the EP for several models. Since the RS expansion is a basic tool in
quantum physics and the Schr\"{o}dinger picture can give us some
quantum-mechanical intuition in QFT, we believe our investigation is
interesting and useful.

Some investigations relevant to the present work can be envisioned. To our
knowledge, scalar field models in Refs.~\cite{12,13,14,15} were seldom
investigated beyond the GA, and so it is worth applying Eq.(27) to those
concrete models as well as to some bosonized models in condensed matter physics.
This paper discussed just ground state solution of Eq.(13), and actually, one
can consider excited state solutions of Eq.(13) to give the EP for the excited
states \cite{8}. Our work here also implies that it is possible to introduce the
optimized procedure to those schemes proposed in Ref.~\cite{2}(i,ii,iii).
Furthermore, since Ref.~\cite{5} has generalized the RS perturbation theory to
the spinor theory, QED and Yang-Mills theory, it should be viable to generalize
the ORSE here to those higher-spin field theories. Besides, since an effective
action contains complete information of a field system, it will be useful to
develop the ORSE here to calculate the effective action \cite{2} (iii). Finally,
Cornwall, Jackiw and Tomboulis developed a generalized EP for composite
operators and calculated it with Rayleigh-Ritz procedure \cite{21}, and it will
be interesting to develop the ORSE here for calculating the generalized EP
which will go beyond the variational result.

\acknowledgments
This project was supported by the Korea Science and Engineering Foundation
through the Center for Strongly correlated materials Research (SNU).
Lu's work was also supported in part by the National
Natural Science Foundation of China under the grant No. 19875034.

\end{document}